\documentstyle[aps,prb,eqsecnum,preprint]{revtex}
\begin{document}
\preprint{cond-mat/9501062}
\draft
\tighten

\title{Conductance fluctuations in a disordered double-barrier junction}
\author{J. A. Melsen and C. W. J. Beenakker}
\address{Instituut-Lorentz, University of Leiden\\
P.O. Box 9506, 2300 RA Leiden, The Netherlands}
\maketitle

\narrowtext
\begin{abstract}We consider the effect of disorder on coherent tunneling
through
two barriers in series, in the regime of overlapping transmission
resonances. We present analytical calculations (using random-matrix theory)
and numerical simulations (on a lattice) to show that strong mode-mixing in the
inter-barrier region induces mesoscopic fluctuations in the conductance $G$ of
universal magnitude $e^2/h$ for a symmetric junction. For an asymmetric
junction, the root-mean-square fluctuations depend on the ratio $\nu$ of the
two tunnel resistances according to $\mbox{rms}\, G = (4e^2/h)\beta^{-1/2}
\nu(1+\nu)^{-2}$, where $\beta = 1 (2)$ in the presence (absence) of
time-reversal
symmetry.
\end{abstract}
\bigskip


\section{Introduction}
Resonant tunneling through two planar barriers in series is a textbook problem
in quantum mechanics. Because of the separation of longitudinal and transverse
motion, the problem is essentially one-dimensional and can be solved in an
elementary way. Realistic double-barrier junctions contain in general some
amount of disorder in the region between the barriers. At low temperatures and
small applied voltages, the inelastic electron-phonon and electron-electron
scattering processes are suppressed, but the elastic scattering by impurities
remains. Scattering events couple the transverse and longitudinal motion of
the tunneling electron, which substantially complicates the
problem but also leads to novel physical effects.

The effects of disorder have been studied in the
past~\cite{FertigEtAl,Leo,Berkovits,LerRai92} with
an emphasis on isolated transmission resonances (energy spacing between the
resonances much greater than their width). Those studies are relevant for
tunneling through a semiconductor quantum well, where the resonances are
widely separated because the barrier separation $L$ is comparable to the
Fermi wavelength $\lambda_{\rm F}$. In the present paper we consider the
opposite regime $L \gg \lambda_{\rm F}$ of strongly overlapping resonances,
relevant to metal structures (where $\lambda_{\rm F}$ is very short,
comparable to the inter-atomic separation), or to tunneling in the plane
of a two-dimensional electron gas (where $L$ can be quite long, because of the
large phase-coherence length).
Two types of disorder can play a role, interface roughness at the barriers and
impurities between the barriers. Interface roughness leads to mesoscopic
(sample-to-sample)
fluctuations in the conductance even in the absence of  any phase coherence,
because
the tunnel probability $\Gamma$ of a single barrier depends strongly on its
thickness. Conductance fluctuations for a single rough tunnel barrier have been
studied by Raikh and Ruzin.~\cite{RaRu} Here we consider the case of impurity
scattering in the absence of interface roughness. Phase coherence is then
essential.

A methodological difference with earlier work
on resonant tunneling is
our use of random-matrix theory to describe the mode-mixing in the
inter-barrier
region.  We assume that the disorder is weak enough that
its effect on the average conductance is negligibly small. This requires a mean
free
path $l \gg \Gamma L$.
Still, the disorder should be sufficiently strong to fully mix the transverse
modes in the inter-barrier region. This requires both $l \ll L/\Gamma$ and $W
\ll L/\Gamma$ (where $W$ is the transverse dimension of the junction). We may
then
describe the disorder-induced mode-mixing
by a random $N\times N$ unitary matrix ($N$ being the
total number of propagating transverse modes at the Fermi energy). This single
assumption permits a complete analytical solution of the statistical properties
of the conductance, using basic results for the so-called circular ensemble
of random matrices.~\cite{Mehta} The circular ensemble is fully characterized
by the symmetry index $\beta$, which equals 1 in the presence of time-reversal
symmetry (circular orthogonal ensemble) and 2 if time-reversal symmetry is
broken by a magnetic field (circular unitary ensemble). (A third possibility,
$\beta = 4$, applies to zero magnetic field in the presence of strong
spin-orbit
scattering.)

As described in Sec. II, we find that the conductance $G$ of the double-barrier
junction exhibits sample-to-sample fluctuations around the classical series
conductance
\begin{equation}
G_{\rm series} = (2e^2/h)N(1/\Gamma_1+1/\Gamma_2)^{-1}.
\end{equation}
(We denote by $\Gamma_1$ and $\Gamma_2$ the transmission probabilities per mode
through barrier 1 and 2, and assume that these are mode-independent and $\ll
1$.)
We find that the root-mean-square fluctuations
$\mbox{rms}\,G$ of the conductance depend only on the ratio $\nu =
\Gamma_1/\Gamma_2$ of the two transmission probabilities, according to
\begin{equation}
\mbox{rms}\, G = \frac{4e^2}{h} \beta^{-1/2}\frac{\nu}{(1+\nu)^2}.
\label{eq:gnu}
\end{equation}
Corrections to Eq.~(\ref{eq:gnu}) are smaller by a factor $e^2/h G_{\rm
series}$,
which is $\ll 1$ if $N\Gamma_i \gg 1$.
For a symmetric junction $(\nu = 1)$ the fluctuations are of order $e^2/h$,
independent of $N$ or $\Gamma_i$ (as long as $N\Gamma_i \gg 1$). This
universality is reminiscent of the
universal conductance fluctuations in diffusive metals.~\cite{Alt85,LeeSt85}
Just as in those systems, we expect the sample-to-sample fluctuations to be
observable in a single sample, as reproducible fluctuations of the conductance
as a function of Fermi energy or magnetic field.

Eq.~(\ref{eq:gnu}) assumes weak disorder, $l \gg \Gamma_iL$ (but still $l\ll
L/\Gamma_i$).
We generalize our results in Sec. III to stronger disorder, when the effects of
the impurities on the average conductance have to be taken into account.
As in a previous paper,~\cite{BeenMel} where we considered a point-contact
geometry, we do this by means of the Dorokhov-Mello-Pereyra-Kumar (DMPK)
equation.~\cite{Dorok82,MPK88}
We find that impurity
scattering leads to the appearance of a weak-localization effect on the
average conductance (observable as a negative magnetoresistance). The
conductance
fluctuations become independent of $\Gamma_1$ and $\Gamma_2$ if $L\gg
l(\Gamma_1^{-1}+\Gamma_2^{-1}).$ A similar conclusion was reached previously by
Iida,
Weidenm\"uller, and Zuk,~\cite{IWZ90} who studied the conductance fluctuations
of a
chain of disordered grains as a function of the coupling strength to two
electron
reservoirs. These authors found that the universal conductance fluctuations are
recovered for a chain length $L$ much greater than some length $L_0$ which is
parametrically greater than the mean free path. A more detailed comparison
with Ref.~\onlinecite{IWZ90} is not possible, because we consider a
homogeneously disordered conductor rather than a chain of disordered grains.

To test our random-matrix description of mode-mixing by weak disorder, we
present
in Sec. IV results from a numerical simulation of a disordered double-barrier
junction defined on a two-dimensional lattice. The agreement with the theory is
quite reasonable.

Two appendices to the paper contain some technical
material which we need in the main text: In App. A we present the analogue of
the Dyson-Mehta formula~\cite{DysonMehta63} for the circular ensemble, which
expresses the variance of the conductance as a Fourier series. In App. B we
discuss the application to our problem of the method of
moments~\cite{Mello,MelloStone} for the DMPK equation.

\section{Double-barrier junction with strong mode-mixing}
The double-barrier junction considered is shown schematically in the inset of
Fig.~1. Since we assume $\lambda_{\rm F} \ll L$,
the scattering matrix $S$ of the whole system can be constructed from the
scattering matrices $S_i$ of the individual barriers.
The $2N\times 2N$ unitary matrix $S_i$ contains two $N \times N$ submatrices
$r_i\vphantom{'}$ and $r_i'$ (reflection from left to left and from right
to right) and two other $N\times N$ submatrices $t_i\vphantom{'}$ and $t_i'$
(transmission from left to right and from right to left). We use the
polar decomposition~\cite{Stoneoverview,MartinLandauer}
\begin{equation}
\label{eq:polar}
S_i = \left(\begin{array}{cc} r_i^{\vphantom{x}} & t_i' \\ t_i^{\vphantom{x}} &
r_i'\end{array}\right) =
\left(\begin{array}{cc} U_i & 0 \\ 0 & V_i \end{array}\right)
\left(\begin{array}{cc} -\mbox{i}(1-\Gamma_i)^{1/2} & \Gamma_i^{1/2} \\
\Gamma_i^{1/2} & -\mbox{i}(1-\Gamma_i)^{1/2} \end{array}\right)
\left(\begin{array}{cc} U_i' & 0 \\ 0 & V_i'\end{array}\right),
\end{equation}
where the $U$'s and $V$'s are $N \times N$ unitary matrices.
In zero magnetic field, $U_i' = U_i^{\rm T}$ and $V_i' = V_i^{\rm T}$,
so that $S_i$ is symmetric --- as it should be in the presence of time-reversal
symmetry.
The transmission matrix $t$ of the whole system is given by
\begin{equation}
\label{eq:tball}
t = t_2^{\vphantom{x}}(1 - r_1' r_2\vphantom{'})^{-1}t_1^{\vphantom{x}}.
\end{equation}
Substitution of the polar decomposition~(\ref{eq:polar}) yields the matrix
product $tt^{\dagger}$ in the form
\begin{mathletters}
\label{eq:ttda}
\begin{eqnarray}
tt^{\dagger} & =  &
V_2^{\vphantom{x}}\left[a+\case{1}{2}b(\Omega+\Omega^{\dagger})\right]^{-1}
V_2^{\dagger},\\
\Omega & = & U_2'V_1^{\vphantom{x}}V_1'U_2^{\vphantom{x}},\\
a & = & [1+(1-\Gamma_1)(1-\Gamma_2)]/\Gamma_1\Gamma_2,\\
b & = & 2\sqrt{(1-\Gamma_1)(1-\Gamma_2)}/\Gamma_1\Gamma_2.
\end{eqnarray}
\end{mathletters}%
The eigenvalues $T_n$ of $tt^{\dagger}$ are related to the eigenvalues
$\exp({\rm i}\phi_n)$ of $\Omega$ by
\begin{equation}
\label{eq:tn}
T_n  = (a + b\cos\phi_n)^{-1}.
\end{equation}
The $T_n$'s determine the conductance $G$ of the double-barrier junction,
according
to the Landauer formula
\begin{equation}
\label{eq:land}
G = G_0\sum_{n=1}^N T_n,
\end{equation}
where $G_0 = 2e^2/h$ is the conductance quantum.

We consider an {\em isotropic} ensemble of double-barrier junctions,
analogous to the isotropic ensemble of disordered wires.\cite{Stoneoverview}
We assume that $l \ll L/\Gamma_i$ and $W \ll L/\Gamma_i$, so that the
tunneling is accompanied by strong mode-mixing: An electron entering
the junction in mode $n$ is randomly distributed among all modes $m$
before leaving the junction. We assume
in this section that mode-mixing is the dominant effect of the disorder, and
that the reduction of the average conductance by the impurity scattering can be
neglected. This requires $l \gg \Gamma_iL$. (The case of stronger disorder is
treated in the next section.) In the polar decomposition~(\ref{eq:polar}) the
mode-mixing is accounted for by the unitary matrices $U$ and $V$. The number of
different unitary matrices is $2\beta$, where $\beta = 1$
in zero magnetic field and $\beta = 2$ if time-reversal symmetry is broken by
a magnetic field.
The isotropic ensemble is the ensemble where the $2\beta$
unitary matrices are {\em independently} and {\em uniformly} distributed
over the unitary group. In other words, the $U$'s and $V$'s are drawn
independently from the circular unitary ensemble (CUE) of random-matrix
theory.~\cite{Mehta}

To determine the statistics of the conductance~(\ref{eq:land}) we need the
probability distribution $P(\{\phi_n\})$ of the eigenvalues of $\Omega$.
For $\beta=2$, $\Omega = U_2'V_1^{\vphantom{x}}V_1'U_2^{\vphantom{x}}$
is the product of  four independent matrices from the CUE, and hence $\Omega$
is also distributed according to the CUE.
For $\beta=1$, $\Omega = U_2^{\rm T}V_1^{\vphantom{x}}V_1^{\rm
T}U_2^{\vphantom{x}}$ is of the form $WW^{\rm T}$ with $W$ a member of  the
CUE. The ensemble
of $\Omega$ is then the circular orthogonal ensemble (COE). The distribution
of the eigenvalues in the CUE and COE is given by~\cite{Mehta}
\begin{equation}
P(\{\phi_n\}) = C\prod_{n < m}\left|
\exp(\mbox{i}\phi_n)-\exp(\mbox{i}\phi_m)\right|^{\beta},
\label{eq:pphi}
\end{equation}
where $C$ is a normalization constant.

We compute the average $\langle A \rangle$ and variance
$\mbox{Var}\,A = \langle A^2\rangle -\langle A\rangle^2$ of linear
statistics $A=\sum_{n=1}^N a(\phi_n)$ on the eigenphases $\phi_n$.
Since in the circular ensemble the $\phi_n$'s are uniformly distributed
in $(0,2\pi)$, the average is exactly equal to
\begin{equation}
\langle A \rangle = \frac{N}{2\pi}\int_0^{2\pi}\mbox{d}\phi\,a(\phi).
\end{equation}
An exact expression for the variance can also be given,~\cite{Mehta} but is
cumbersome to evaluate. For $N\gg 1$ we can use a variation on the Dyson-Mehta
formula~\cite{DysonMehta63} (derived in App.\ A),~\cite{Forrester}
\begin{mathletters}
\label{eq:vara}
\begin{eqnarray}
\mbox{Var}\,A & = & \frac{1}{\pi^2\beta}\sum_{n=1}^{\infty} n|a_n|^2 +{\cal
O}(N^{-1}),\\
a_n & = & \int_0^{2\pi}\mbox{d}\phi\, \mbox{e}^{{\rm i}n\phi}a(\phi).
\end{eqnarray}
\end{mathletters}%
For the conductance [given by Eqs.~(\ref{eq:tn}) and (\ref{eq:land})], we
substitute $a(\phi) = (a + b\cos\phi)^{-1}$, with Fourier coefficients
$a_n = {2\pi}(a^2-b^2)^{-1/2}b^{-n}\left[(a^2-b^2)^{1/2}-a\right]^n$. The
results are
\begin{eqnarray}
\label{eq:mean}
\langle G/G_0 \rangle & = & N(1/\Gamma_1+1/\Gamma_2-1)^{-1},\\
\label{eq:var}
\mbox{Var}\, G/G_0 & = &
\frac{4}{\beta}\frac{(1-\Gamma_1)(1-\Gamma_2)\Gamma_1^2\Gamma_2^2}
{(\Gamma_1+\Gamma_2-\Gamma_1\Gamma_2)^4}.
\end{eqnarray}

Equation~(\ref{eq:mean}) for the average conductance is what one would expect
from
classical addition of the resistances $(N\Gamma_i G_0)^{-1}$ of the individual
barriers. (The $-1$ in Eq.~(\ref{eq:mean}) corrects for a double counting of
the contact resistance and becomes irrelevant for $\Gamma_i \ll 1$.) Each
member of the ensemble contains a different
set of overlapping transmission resonances, and the ensemble average removes
any trace of resonant tunneling in $\langle G\rangle$. In a previous
paper,~\cite{MelBeen} we have shown that the average conductance differs
drastically from
the series conductance if the double-barrier junction is connected to a
superconductor, but here we consider only normal-metal conductors.

Eq.~(\ref{eq:var}) for the conductance fluctuations tells us
that $\mbox{Var}\,G$ becomes completely independent of $N$ in the
limit $N\rightarrow \infty$.
[More precisely, corrections to Eq.~(\ref{eq:var}) are of order $\langle G/G_0
\rangle^{-1}$, which is $\ll 1$ if $N\Gamma_i \gg 1$.]
Since $\Gamma_i \ll 1$, we may simplify Eq.~(\ref{eq:var}) to
\begin{equation}
\mbox{Var}\,G/G_0 = \frac{4}{\beta}
\frac{\Gamma_1^2\Gamma_2^2}{(\Gamma_1+\Gamma_2)^4},
\end{equation}
which depends only on the ratio $\Gamma_1/\Gamma_2$ and not on the individual
$\Gamma_i$'s.
The variance reaches a $\Gamma$-independent maximum for two equal
barriers,
\begin{equation}
\mbox{Var}\,G/G_0 = \case{1}{4}\beta^{-1},\ \ \ \ \mbox{if}\ \ \Gamma_1 =
\Gamma_2.
\label{eq:varlimit}
\end{equation}
The variance is almost twice the result $\case{2}{15}\beta^{-1}$ for an
isotropic
ensemble of disordered wires,~\cite{Mello,MelloStone} and precisely twice the
result $\case{1}{8}\beta^{-1}$ for an isotropic ensemble of ballistic quantum
dots.~\cite{IWZ90,Jalabert,BarangerMello}

\section{Effects of strong disorder}
In this section we relax the assumption $l \gg \Gamma_i L$ of Sec.~II, to
include the case that the impurity scattering is sufficiently strong to
affect the average conductance. We assume $W \ll L$, so that
we are justified in using an isotropic distribution for the scattering
matrix $S_L$ of the inter-barrier region.~\cite{Stoneoverview} The scattering
matrix $S$ of the entire system is now composed from the three scattering
matrices $S_1, S_L,$ and $S_2$ in series. The composition is most easily
carried out in terms of the transfer matrices $M_1, M_L$, and $M_2$ associated
with  $S_1, S_L$, and $S_2$, respectively. The transfer matrix $M$ of the
entire system is the matrix product $M = M_2M_LM_1$, so the total
distribution $P(M)$ is a convolution of the individual distributions $P_1(M_1),
P_L(M_L),$ and $P_2(M_2)$: $P = P_2*P_L*P_1$, where the convolution $*$ is
defined
by
\begin{equation}
P_i*P_j(M) = \int \mbox{d}M'\,P_i(MM'^{-1})P_j(M').
\end{equation}
The isotropy assumption implies that each distribution $P_i(M_i)$ is only a
function of the eigenvalues of $M_i^{\vphantom{\dagger}}M_i^{\dagger}$.

We now use the fact that the convolution of isotropic distributions of
transfer matrices  commutes. (A proof is given in Ref.~\onlinecite{BeenMel}.)
This permits us to consider an equivalent system, with transfer matrix $M =
M_LM_2M_1$, where all disorder is at one side of the double-barrier junction
---
instead of in between the barriers. The $L$-dependence of the distribution of
transmission eigenvalues for this system is governed by the DMPK
equation,~\cite{Dorok82,MPK88}
\begin{mathletters}
\label{eq:dmpk}
\begin{eqnarray}
\frac{\partial}{\partial s}  P(\{\lambda_n\},s) & = & \frac{2}{\beta N
+2-\beta} \sum_{i=1}^N\frac{\partial}{\partial
\lambda_i}\left(\lambda_i(1+\lambda_i)
J\frac{\partial}{\partial \lambda_i}\frac{P}{J}\right),\\
J & = & \prod_{i<j}|\lambda_i-\lambda_j|^{\beta},
\end{eqnarray}
\end{mathletters}%
where $s=L/l$ and $\lambda_n = (1-T_n)/T_n$. The initial condition ($s
\rightarrow 0$) of Eq.~(\ref{eq:dmpk}) corresponds to taking $M_L=1$, which
implies
for $P$ the isotropic ensemble given by Eq.~(\ref{eq:pphi}).

To compute the $L$-dependence of the mean and variance of the conductance, we
use the method of moments of
Mello and Stone,~\cite{Mello,MelloStone} who have derived a hierarchy of
differential equations for the moments of ${\cal T}_q \equiv
\sum_{n=1}^NT_n^q$.
The hierarchy closes order by order in an expansion in powers of $1/N$. Mello
and Stone considered a ballistic initial condition, corresponding to
$\langle {\cal T}_q^p\rangle \rightarrow N^p$ for $s\rightarrow 0$. We have the
different initial condition of a double-barrier junction. The differential
equations and initial conditions for the moments are given in App.\ B.
For the mean conductance and its variance we obtain
\begin{eqnarray}
\langle G/G_0 \rangle & = &\frac{N}{s+\rho} +\frac{1}{3}(1-2/\beta)-
\frac{1-2/\beta}{(s+\rho)^3} \left(s^2 -
s(a-\rho-\rho^2)+\case{1}{3}\rho^3\right),\label{eq:gs}\\
\mbox{Var}\,G/G_0 & = & \frac{2}{15\beta} +
\frac{2}{\beta(s+\rho)^6}\left(s^2(\case{1}{2}a^2+\case{1}{2}\rho^2-
2a\rho^2+\rho^4)\right.\nonumber\\
 & & + \left.s (-{2a^2}\rho+{2a}\rho^3-\case{2}{5}\rho^5)
+\case{1}{2}a^2\rho^2-\case{1}{2}\rho^4-\case{1}{15}\rho^6\right)
\label{eq:vars},
\end{eqnarray}
where $a$ has been defined in Eq.~(\ref{eq:ttda}) and $\rho$ is defined by
\begin{equation}
\rho= 1/\Gamma_1+1/\Gamma_2-1.
\end{equation}

Corrections to Eqs.~(\ref{eq:gs}) and (\ref{eq:vars}) are of order
$(s+\rho)/N$.
For two equal barriers ($\Gamma_1 = \Gamma_2 \equiv \Gamma$) in the
limit $\Gamma \rightarrow 0$ at fixed $\Gamma s$,
Eqs.~(\ref{eq:gs}) and (\ref{eq:vars}) simplify to
\begin{eqnarray}
\delta G/G_0 & \equiv & \langle G/G_0\rangle-N(s+\rho)^{-1}=
\frac{1}{3}(1-2/\beta)-\frac{1-2/\beta}{(2+\Gamma
s)^3}\left(\frac{8}{3}+2\Gamma s\right),\label{eq:gsimp}\\
\mbox{Var}\,G/G_0 & = & \frac{2}{15\beta} + \frac{4}{\beta(2+\Gamma s)^6}\left(
\Gamma^2s^2 + \frac{8}{5}\Gamma s + \frac{28}{15}\right).\label{eq:varsimp}
\end{eqnarray}
Eqs. (\ref{eq:gsimp}) and (\ref{eq:varsimp}) are plotted in Fig.\ 1 (for $\beta
=1$).
In the limit of large disorder
($\Gamma s \gg 1$), we recover the familiar results\cite{Mello,MelloStone}
 for a disordered wire: $\delta G/G_0 = \case{1}{3}(1-2/\beta)$,
$\mbox{Var}\,G/G_0 = \case{2}{15}\beta^{-1}$ (indicated by arrows in Fig.\ 1).
In the opposite limit $\Gamma s \ll 1$, we find $\delta G = 0$, $\mbox{Var}\,
G/G_0 = \case{1}{4}\beta^{-1}$ --- as in Sec.~II [cf. Eqs.~(\ref{eq:mean}) and
(\ref{eq:varlimit})].

\section{Numerical simulations}
To test our results we have performed numerical simulations, using the
recursive Green's function method of Ref.~\onlinecite{Bar91}.
The disordered inter-barrier region was modeled by a tight-binding
Hamiltonian on a two-dimensional square lattice with lattice constant {\em d}.
The Fermi energy was chosen at 1.5\,$u_0$ from the band bottom,
with $u_0 = \hbar^2/2 m d^2$.
Disorder was introduced by randomly assigning a value between $\pm
\case{1}{2}U_{\rm D}$ to the on-site potential of the lattice points in a
rectangle
with $L = 142d$, $W=71d$ (corresponding to $N=30$).
We chose $U_{\rm D} = 0.6\,u_0$, corresponding to $L/l = 0.9$.
The transfer matrix $M_L$ was computed numerically, and then multiplied
with the transfer matrices $M_1$ and $M_2$ of the two barriers (which we
constructed analytically, given the mode-independent tunnel probabilities
$\Gamma_1$
and $\Gamma_2$).
We took $\Gamma_2 = 0.15$ and  varied $\Gamma_1$ between
$0.05$ and $0.5$. These parameter values were chosen in order to be
close to the regime $\Gamma_i L \ll l \ll L/\Gamma_i$, $W \ll L/\Gamma_i$ in
which disorder is expected to cause strong mode-mixing, without having a
large effect on the average conductance (the regime studied in Sec. II).

In Fig.\ 2 we show the comparison between theory and simulation. The
solid curve is $\mbox{Var}\, G/G_0$ computed from 2250 realizations of
the disorder potential. The dotted curve is the theoretical prediction
from Eq.~(\ref{eq:vars}) for the parameter values of the simulation (and
for $\beta = 1$, since there was no magnetic field). There are no
adjustable parameters. The agreement is quite reasonable. It is likely
that the remaining discrepancy is due to the fact that the theoretical
condition $N\Gamma_i \gg 1$ was not well met in the simulation (where
$N\Gamma_2 = 4.5$). The value $N=30$ of the simulation is already at
the limit of our computational capabilities and we are not able to
provide a more stringent numerical test of the theory.

\acknowledgements
Discussions with P. W. Brouwer have been most helpful.
This research was supported by the ``Ne\-der\-land\-se or\-ga\-ni\-sa\-tie voor
We\-ten\-schap\-pe\-lijk
On\-der\-zoek'' (NWO) and by the ``Stich\-ting voor Fun\-da\-men\-teel
On\-der\-zoek der Ma\-te\-rie'' (FOM).

\appendix
\section{Dyson-Mehta formula for the circular ensemble}
The variance $\mbox{Var}\,A$ of a linear statistic $A = \sum_{n=1}^N a(\phi_n)$
on the eigenphases is given by a double integral,
\begin{equation}
\mbox{Var}\, A =
-\int_0^{2\pi}\mbox{d}\phi\int_0^{2\pi}\mbox{d}\phi'\,a(\phi)a(\phi')
K(\phi,\phi'),\label{eq:vark}
\end{equation}
over the two-point correlation function
\begin{equation}
K(\phi,\phi') = \langle
\rho(\phi)\rangle\langle\rho(\phi')\rangle-\langle\rho(\phi)\rho(\phi')\rangle.
\end{equation}
The brackets $\langle \cdots \rangle$ denote an average over the circular
ensemble, and
\begin{equation}
\rho(\phi) = \sum_{n=1}^N \delta(\phi-\phi_n)
\end{equation}
is the microscopic density of eigenphases.
In this appendix we compute $K(\phi,\phi')$ in the large $N$-limit, using the
method of functional derivatives of Ref.~\onlinecite{Been93}. This leads to
Eq.~(\ref{eq:vara}) for $\mbox{Var}\,A$, which is the analogue for the circular
ensemble of the Dyson-Mehta formula for the Gaussian
ensemble.~\cite{DysonMehta63} The analogy is straightforward, but we have not
found it in the literature.~\cite{Forrester}

We consider a generalized circular ensemble, with probability distribution
\begin{mathletters}
\begin{eqnarray}
P_V(\{\phi_n\}) & = & C\, \mbox{exp}\left[
-\beta\left(\sum_{i<j}U(\phi_i-\phi_j)+\sum_{i=1}^NV(\phi_i)\right)\right],\\
C^{-1} & = &
\int_0^{2\pi}\mbox{d}\phi_1\,\int_0^{2\pi}\mbox{d}\phi_2\,\cdots
\int_0^{2\pi}\mbox{d}\phi_N\, P_V(\{\phi_n\}),\\
U(\phi) & = & - \ln|2\sin\case{1}{2}\phi|.
\end{eqnarray}
\label{eq:pv}
\end{mathletters}%
The ``potential'' $V(\phi)$ is arbitrary. If $V \equiv 0$, Eq.~(\ref{eq:pv}) is
the same as the distribution~(\ref{eq:pphi}) of the circular ensemble. The
brackets $\langle \cdots\rangle_V$ denote an average with the $V$-dependent
distribution~(\ref{eq:pv}). Following Ref.~\onlinecite{Been93}, we express the
two-point correlation function as a functional derivative of the density
with respect to the potential,
\begin{equation}
K(\phi,\phi')  = \frac{1}{\beta}\frac{\delta \langle
\rho(\phi)\rangle_V}{\delta V(\phi')}.\label{eq:kfd}
\end{equation}
The functional derivative can be computed in the large-$N$ limit from the
relationship\cite{Dyson}
\begin{equation}
-\int_0^{2\pi} \mbox{d}\phi'\,U(\phi-\phi')\langle \rho(\phi')\rangle_V =
V(\phi)+\mbox{const}.
\label{eq:fequi}
\end{equation}
Corrections to Eq.~(\ref{eq:fequi}) are smaller by a factor $1/N$.
The additive constant is obtained from the normalization $\int \mbox{d}\phi\,
\langle\rho(\phi)\rangle_V = N$.

Fourier transformation of Eq.~(\ref{eq:fequi}) yields
\begin{equation}
-\frac{\pi}{|n|}\langle \rho_n\rangle_V = V_n,\ \ n \neq 0.
\end{equation}
We have defined the Fourier coefficients
\begin{equation}
f_n = \int_0^{2\pi}\mbox{d}\phi\, \mbox{e}^{{\rm i}n\phi} f(\phi),
\end{equation}
and we have used that $U_n = \pi/|n|$ for $n\neq 0$.
{}From Eqs.~(\ref{eq:kfd}) and~(\ref{eq:fequi})
we see that  $K(\phi,\phi')=K(\phi-\phi')$ depends
on the difference $\phi-\phi'$ only, and is independent of $V$. The Fourier
coefficients of $K(\phi)$ are
\begin{mathletters}
\label{eq:kp}
\begin{equation}
\label{eq:kn}
K_n = -|n|/\pi\beta,
\end{equation}
for $n \neq 0$.
Since $K_0 = 0$ by definition, Eq.~(\ref{eq:kn}) holds in fact for all $n$.
Inversion of the Fourier transform yields the correlation function
\begin{equation}
K(\phi)=-\frac{1}{\pi^2\beta}\frac{\mbox{d}^2}{\mbox{d}\phi^2}\ln\left|\sin
\case{1}{2}\phi\right|,
\end{equation}
\end{mathletters}%
which has an integrable singularity at $\phi = 0$. For $\phi \neq 0$, $K(\phi)
= [4\pi^2\beta \sin^2(\phi/2)]^{-1}$. Substitution of Eq.~(\ref{eq:kp}) into
Eq.~(\ref{eq:vark}) gives the required analogue of the Dyson-Mehta formula for
the large $N$-limit of the variance of a linear statistic,
\begin{eqnarray}
\mbox{Var}\, A & = &
-\frac{1}{\pi^2\beta}\int_{0}^{2\pi}\mbox{d}\phi\,\int_0^{2\pi}\mbox{d}
\phi'\,\left(\frac{\mbox{d}a(\phi)}{\mbox{d}\phi}\right)\left(\frac{\mbox{d}
a(\phi')}{\mbox{d}\phi'}\right)\ln\left|\sin\frac{\phi-\phi'}{2}\right|
\nonumber\\
 & = & \frac{1}{\pi^2\beta}\sum_{n=1}^{\infty}n|a_n|^2.
\label{eq:varDysum}
\end{eqnarray}

\section{Moment expansion of the DMPK equation}
Mello and Stone~\cite{MelloStone} have derived from the DMPK
equation~(\ref{eq:dmpk}) a hierarchy of differential equations for the moments
of ${\cal T}_q = \sum_{n=1}^N T_n^q$. The hierarchy closes order by order in
the series expansion
\begin{mathletters}
\label{eq:momdefs}
\begin{eqnarray}
\langle {\cal T}^p \rangle & = & N^p f_{p,0}(s) + N^{p-1} f_{p,1}(s) + N^{p-2}
f_{p,2}(s)
+ \ldots ,\\
\langle {\cal T}^p {\cal T}_2 \rangle & =  & N^{p+1} g_{p+1,0}(s) + N^{p}
g_{p+1,1}(s) + N^{p-1} g_{p+1,2}(s) + \ldots ,\\
\langle {\cal T}^p {\cal T}_3 \rangle & =  & N^{p+1} h_{p+1,0}(s) + N^{p}
h_{p+1,1}(s) + N^{p-1} h_{p+1,2}(s) + \ldots ,\\
\langle {\cal T}_{\vphantom{x}}^p {\cal T}_2^2 \rangle & =  & N^{p+2}
l_{p+2,0}(s) + N^{p+1} l_{p+2,1}(s) + N^{p} l_{p+2,2}(s) + \ldots ,
\end{eqnarray}
\end{mathletters}%
where we have defined ${\cal T} \equiv {\cal T}_1$.
For a calculation of $\mbox{Var}\, G$
we need to determine $\langle {\cal T}^p \rangle$ down to ${\cal O}(N^{p-2})$,
$\langle {\cal T}^p{\cal T}_2\rangle$ down to ${\cal O}(N^{p})$, and $\langle
{\cal T}^p{\cal T}_3\rangle$
and $\langle {\cal T}_{\vphantom{x}}^p{\cal T}_2^2\rangle$ only to the highest
occurring order.
The resulting set of differential equations we have to solve
is~\cite{MelloStone}
\begin{mathletters}
\label{eq:feqs}
\begin{eqnarray}
&& f'_{p,0}(s)  +  pf_{p+1,0}(s) = 0,\\
&& g'_{p,0}(s)  +  (p+3)g_{p+1,0}(s) = 2f_{p+1,0}(s),\\
&& f'_{p,1}(s)  +  pf_{p+1,1}(s) =
(1-2/\beta)\left[f'_{p,0}(s)+pg_{p,0}(s)\right],\\
&& l'_{p,0}(s)  +  (p+6)l_{p+1,0}(s) = 4g_{p+1,0}(s),\\
&& h'_{p,0}(s)  + (p+5)h_{p+1,0}(s) = 6g_{p+1,0}(s)-3l_{p+1,0}(s),\\
&& g'_{p,1}(s)  +  (p+3)g_{p+1,1}(s) = 2f_{p+1,1}(s)-(1-2/\beta)\left[
-g'_{p,0}(s)+2g_{p,0}(s)\right.\nonumber\\
&& \hspace{5cm}\left.-4h_{p,0}(s)-(p-1)l_{p,0}(s)\vphantom{g'}\right],\\
&& f'_{p,2}(s)  +  pf_{p+1,2}(s) =
(1-2/\beta)\left[f'_{p,1}(s)+pg_{p,1}(s)\right]\nonumber\\
&& \hspace{4cm}+2\beta^{-1}p(p-1)\left[g_{p-1,0}(s)-h_{p-1,0}(s)\right].
\end{eqnarray}
\end{mathletters}%

We need to determine the initial conditions $f(0), g(0), h(0),$ and $l(0)$
from the distribution function (\ref{eq:pphi}) for the eigenphases in
the circular ensemble. In the large-$N$ limit, the
linear statistic ${\cal T}_q$ on the eigenphases has a Gaussian distribution
with
a width of order $N^0$.
Therefore, if we write
${\cal T}_q = \langle{\cal T}_q\rangle + \delta {\cal T}_q$,
we know that $\langle {\cal T}_q \rangle = {\cal O}(N)$,
$\langle \delta {\cal T}_q \rangle = 0$,
$\langle (\delta {\cal T}_q)^{2n+1} \rangle = {\cal O}(N^{-1})$ and
$\langle (\delta {\cal T}_q)^{2n}\rangle = {\cal O}(N^0)$.
This implies that, for $s \rightarrow 0$,
\begin{mathletters}
\begin{eqnarray}
\langle {\cal T}^p \rangle & = & \langle {\cal T} \rangle^p +
\case{1}{2}p(p-1)\langle {\cal T} \rangle^{p-2}\langle (\delta {\cal
T})^2\rangle + {\cal O}(N^{p-4}),\\
\langle {\cal T}^p {\cal T}_2 \rangle & = & \langle {\cal T} \rangle^p\langle
{\cal T}_2\rangle + {\cal O}(N^{p-1}),\\
\langle {\cal T}^p {\cal T}_3 \rangle & =  & \langle {\cal T} \rangle^p\langle
{\cal T}_3\rangle + {\cal O}(N^{p-1}),\\
\langle {\cal T}^p {\cal T}_2^2 \rangle & = & \langle {\cal T} \rangle^p\langle
{\cal T}_2\rangle^2 + {\cal O}(N^p).
\end{eqnarray}
\label{eq:texp}
\end{mathletters}%
The average $\langle (\delta {\cal T})^2\rangle $ is just $\mbox{Var}\,G/G_0$,
which is given by Eq.~(\ref{eq:var}),
\begin{equation}
\langle (\delta {\cal T})^2\rangle = \beta^{-1}b^2\rho^{-4}+{\cal O}(N^{-1}).
\end{equation}
The other averages in Eq.~(\ref{eq:texp}) follow from
\begin{equation}
\langle {\cal T}_q \rangle  =
\frac{N}{2\pi}\int_0^{2\pi}\mbox{d}\phi\,(a+b\cos\phi)^{-q}.
\end{equation}
The resulting initial conditions read
\begin{mathletters}
\begin{eqnarray}
f_{p,0}(0) & = & \rho^{-p},\ \ f_{p,1}(0) = 0,\ \ f_{p,2}(0) =
\case{1}{2}\beta^{-1}p(p-1)\rho^{-(p+2)} b^2,\\
g_{p,0}(0) & = & a\rho^{-(p+2)},\ \ g_{p,1}(0) = 0,\\
h_{p,0}(0) & = & \rho^{-(p+2)}(\case{3}{2}a^2\rho^{-2}-\case{1}{2}),\ \
l_{p,0}(0) = a^2\rho^{-(p+4)}.
\end{eqnarray}
\end{mathletters}%

The set of differential equations~(\ref{eq:feqs}) can be solved by substitution
of the following Ansatz for the $p$-dependence (adapted from
Ref.~\onlinecite{Marc}):
\begin{equation}
x_{p,l}(s)=(s+\rho)^{-(p+2l+n)}[p^2\varphi(s)+p\chi(s)+\psi(s)],
\end{equation}
where $n=0$ if $x$ is $f$, $n=3$ if $x$ is $g$, and $n=6$ if $x$ is $h$ or $x$
is $l$.
The mean and variance of the conductance, to order $N^{-1}$, then follow
from
\begin{eqnarray}
\langle G/G_0 \rangle  & = & Nf_{1,0}(s) + f_{1,1}(s),\\
\mbox{Var}\,G/G_0 & = & N^2[f_{2,0}(s)-f_{1,0}(s)^2] + N[f_{2,1}(s)
-2f_{1,0}(s)f_{1,1}(s)] + \nonumber\\
& & f_{2,2}(s)-2f_{1,0}(s)f_{1,2}(s)-f_{1,1}(s)^2.
\end{eqnarray}
The results are Eqs.~(\ref{eq:gs}) and~(\ref{eq:vars}).

\begin{figure}[tb]
\hspace*{\fill}
\hspace*{\fill}
\caption[]{Weak-localization correction $\delta G$ to the average
conductance (in units of $G_0 = 2e^2/h$) and root-mean-square fluctuations
$\mbox{rms}\, G \equiv (\mbox{Var}\, G)^{1/2}$, computed from
Eqs.~(\ref{eq:gsimp}) and (\ref{eq:varsimp}) for $\beta = 1$. The arrows give
the limit $\Gamma L/l \gg 1$. The inset shows
the geometry of the double-barrier junction (the disordered region is dotted).
The curves plotted in the
figure are for a symmetric junction, $\Gamma_1 = \Gamma_2 \equiv \Gamma \ll 1$.
\label{fig1}}
\end{figure}

\begin{figure}[tb]
\hspace*{\fill}
\hspace*{\fill}
\caption[]{Solid curve: variance of the conductance from a numerical
simulation of an ensemble of disordered double-barrier junctions ($L/W=2$,
$N=30$, $s = 0.9$), as a function of the ratio $\Gamma_1/\Gamma_2$,
with $\Gamma_2=0.15$ held constant. There is no magnetic field
($\beta=1$). The dashed curve is the prediction from
Eq.~(\protect\ref{eq:vars}). There are no adjustable parameters.
\label{fig2}}
\end{figure}

\end{document}